\documentclass{PoS}

\usepackage{ amssymb }

\title{The Strangeness and Charmness of Nucleon from Overlap Fermions}

\ShortTitle{The Strangeness and Charmness of Nucleon From Overlap Fermions}

\author{\speaker{M. Gong}\\
        Dept. of Physics and Astronomy, University of Kentucky, Lexington, KY 40506, USA\\
        E-mail: \email{gongming@pa.uky.edu}}

\author{A. Li\\
        Dept. of Physics, Duke University, Durham, NC 27708, USA\\
        E-mail: \email{anyili@phy.duke.edu}}

\author{A. Alexandru\\
        Dept. of Physics, George Washington University, Washington, DC 20052, USA\\
        E-mail: \email{aalexan@gwu.edu}}

\author{T. Draper\\
        Dept. of Physics and Astronomy, University of Kentucky, Lexington, KY 40506, USA\\
        E-mail: \email{draper@pa.uky.edu}}

\author{K.F. Liu\\
        Dept. of Physics and Astronomy, University of Kentucky, Lexington, KY 40506, USA\\
        E-mail: \email{liu@pa.uky.edu}
\begin{center} ($\chi$QCD Collaboration) \end{center}}

\abstract{The calculation of the strangeness and charmness of the nucleon is presented with overlap fermion action on $2+1$ flavor domain wall fermion configurations.  
We adopt stochastic grid sources and the low mode substitution  technique to improve the signals of nucleon correlation functions and the loops. 
The calculation is done on a $24^3\times 64$ lattice with $m_l=0.005$, $m_h=0.04$, and $a^{-1}=1.73\,{\rm GeV}$.
We find $ f_{T_{s}} = 0.048(15)$ and $f_{T_{c}} = 0.029(43)$.
}

\FullConference{The XXIX International Symposium on Lattice Field Theory, Lattice2011\\
July 10-16, 2011, Squaw Valley, Lake Tahoe, \\
California, USA
}

\begin{document}

\section{Introduction}

The strangeness and charmness contents of the nucleon are interesting aspects of nucleon structure.
Despite the difficulty in measuring these quantities experimentally, the strangeness of nucleon still attracts much interest because it is believed to make a large contribution to the cross section between some dark matter candidates and ordinary matter~\cite{darkmatter}.

The strangeness and charmness of the nucleon can be defined as dimensionless fractions:
\begin{eqnarray}
f_{T_{s}} &= \frac{m_{s} \left< N | \bar{s} s | N \right>}{m_N}\\
f_{T_{c}} &= \frac{m_{c} \left< N | \bar{c} c | N \right>}{m_N}
\end{eqnarray}
There are lattice QCD calculations with dynamical fermions by several groups~\cite{jlqcd,milc,qcdsf,Engelhardt,Young}.
Two different methods are used.
One is to calculate $\left< N | \bar{q} q | N \right> = \frac{\partial m_N}{\partial m_q}$ from the Feynman-Hellmann theorem.
Another is to calculate the disconnected three-point functions
\begin{equation}
R(t^\prime, t, t_0) = \frac{< N(t) \bar{q}q(t^\prime) \bar{N}(t_0) > - < N(t) \bar{N}(t_0) > < \bar{q}q(t^\prime)>}{< N(t) \bar{N}(t_0) >}
\end{equation}
where $\bar{q}q(t^\prime)$ and the nucleon sink $N(t)$ are zero-momentum operators.
We adopt the latter method to calculate the strangeness and charmness of the nucleon.
To extract the form factor from the ratio of the three-point function and the two-point function, we use a summed ratio to enhance the signal and suppress excite-state contamination~\cite{sumratio}
\begin{equation}
R^\prime(t, t_0) = \sum_{t^\prime=t_0+1}^{t-1} R(t^\prime, t, t_0)
\end{equation}
and do a linear fit for large $t$:
\begin{equation}
R^\prime(t, t_0) {}_{\stackrel{\longrightarrow}{t \gg t_0}} const. + t \left< N | \bar{s} s | N \right>
\end{equation}

\section{Strategies for the computation}

\subsection{Overlap valence fermion on domain-wall sea}

The overlap fermion action obeys chiral symmetry at finite lattice spacing and is free of $O(a)$ errors. 
It is shown that the effective quark propagator of the massive overlap fermion has the same form as that of the continuum~\cite{continuum}.
The $O(m^2a^2)$ error, which is important in the charm region, is estimated to be small on quenched lattices~\cite{quench1,dong07} and even 
smaller on the dynamical domain-wall sea~\cite{dong09} due to HYP smearing.
Thus, it is shown that it can be used for both charm and light quarks on the $24^3 \times 64$ lattice DWF configurations~\cite{li10}.

We use the valence overlap fermion on $N_f = 2+1$ domain wall dynamical configurations obtained from the RBC and UKQCD collaborations~\cite{rbc}.
This is a mixed action approach.
Since the valence is a chiral fermion, only one extra low-energy constant $\Delta_{mix}$ needs to be determined~\cite{deltamix1,deltamix2}, which turns out to be small~\cite{li10}.

\subsection{The stochastic grid source}

We introduce $Z_3$ and $Z_4$ stochastic grid sources to gain more efficiency.
A grid source is defined as
\begin{equation}
\eta(\vec{x}) = \sum_{\vec{i}\in \mathcal{G}} \theta_i \delta_{\vec{x},\vec{i}}
\end{equation}
where $\mathcal{G}$ is a sparse regular grid of lattice sites on time slice $t=0$, and $\theta_i$ is the random phase on site $\vec{i}$.
The corresponding quark propagator is
\begin{eqnarray}
G(\vec{y},\eta) &=& D^{-1}(\vec{y},\vec{x}) \eta(\vec{x}) \nonumber\\
       &=& \sum_{\vec{i}\in \mathcal{G}} \theta_i D^{-1}(\vec{y},\vec{x}) \delta_{\vec{x},\vec{i}} = \sum_{\vec{i}\in \mathcal{G}} \theta_i G(\vec{y},\vec{i})
\end{eqnarray}
And the loop on site $\vec{i}$ is:
\begin{equation}
L(\vec{i}) = \theta^\dag_i G(\vec{i},\eta) \rightarrow G(\vec{i},\vec{i})
\end{equation}

\subsection{The low mode substitution technique for the proton correlation functions}

The proton correlation function is
\begin{eqnarray}
C_{proton}(\vec{y}, \vec{x}; \Gamma; G^{(u)}, G^{(d)}, G^{(u^\prime)}) = \epsilon^{abc} \epsilon^{a^\prime b^\prime c^\prime} \left[ tr\left( \Gamma G^{(u)aa^\prime}(\vec{y}, \vec{x}) \underline{G}^{(d)bb^\prime}(\vec{y}, \vec{x}) G^{(u^\prime)cc^\prime}(\vec{y}, \vec{x})\right) \right. \nonumber\\
\left.+ tr\left( \Gamma G^{(u)aa^\prime}(\vec{y}, \vec{x}) \right) tr\left( \underline{G}^{(d)bb^\prime}(\vec{y}, \vec{x}) G^{(u^\prime)cc^\prime}(\vec{y}, \vec{x})\right) \right]
\end{eqnarray}
where the $G^{(u/d)aa^\prime}(\vec{y}, \vec{x})$ is the $u$/$d$ quark propagator from $\vec{x}$ to $\vec{y}$ with color indices $a$,$a^\prime$ and the $\underline{G}$ is defined~\cite{proton} as $(\tilde{C}G\tilde{C}^{-1})^T$ with the charge conjugation operator $C$ and $\tilde{C} = C \gamma_5 = \gamma_2 \gamma_4 \gamma_5$.
The trace and the transpose operations are defined on Dirac space.

With the masses of the $u$ and the $d$ quarks set to the same value, $G^{(d)}$, $G^{(u)}$, and $G^{(u')}$ are the same propagator.  The correlation function $C(G_1,G_2,G_3)$ is a linear function of each of its three arguments.

With the $Z_3$ grid source, the quark propagator is the sum of the propagators from different sites on the grid with different $Z_3$ phases
\begin{equation}
G = \sum_{i \in \mathcal{G}} \theta_i G_i
\end{equation}
and the proton correlation function is
\begin{eqnarray}
C(G, G, G) =& C(\sum_i \theta_i G_i, \sum_j \theta_j G_j, \sum_k \theta_k G_k) \nonumber \\
           =& \sum_{i,j,k} \theta_i \theta_j \theta_k C(G_i, G_j, G_k) \nonumber \\
           \rightarrow&  \sum_{i,j,k} \delta_{i,j,k} C(G_i, G_j, G_k) \nonumber \\
           \rightarrow&  \sum_{i} C(G_i, G_i, G_i)
\end{eqnarray}
The limit is at that with infinite noise sources or infinite configurations. Otherwise, there is unbiased contamination by the propagators from different sites.

With deflation in the quark matrix inversion, we have the propagator with low-mode and high-mode parts as well:
\begin{eqnarray}
G =& G^H + G^L \nonumber\\
  =& \sum_{i}\theta_i (G^H_i + G^L_i) \nonumber \\
  =& G^H + \sum_{i}\theta_i G^L_i
\end{eqnarray}
and the proton correlation function is
\begin{eqnarray}
C(G, G, G) =& C(G^H + \sum_{i}\theta_i G^L_i, G^H + \sum_{j}\theta_j G^L_j, G^H + \sum_{k}\theta_k G^L_k) \nonumber \\
           =& C(G^H, G^H, G^H) + \sum_{i} C(\theta_i G^L_i, \theta_i G^L_i, \theta_i G^L_i) \nonumber \\
           &+ \sum_{i} C(\theta_i G^L_i, G^H, G^H) + \sum_{i} C(G^H, \theta_i G^L_i, G^H) + \sum_{i} C(G^H, G^H, \theta_i G^L_i) \nonumber \\
           &+ \sum_{i} C(\theta_i G^L_i, \theta_i G^L_i, G^H) + \sum_{i} C(\theta_i G^L_i, G^H, \theta_i G^L_i) + \sum_{i} C(G^H, \theta_i G^L_i, \theta_i G^L_i) \nonumber \\
           &+ \sum_{i \neq j} C(\theta_i G^L_i, \theta_j G^L_j, G^H) + \sum_{i \neq j} C(\theta_i G^L_i, G^H, \theta_j G^L_j) + \sum_{i \neq j} C(G^H, \theta_i G^L_i, \theta_j G^L_j) \nonumber \\
           &+ \sum_{i \neq j \, or \, j \neq k \, or \, k \neq i} C(\theta_i G^L_i, \theta_j G^L_j, \theta_k G^L_k)
\end{eqnarray}

By design, the three quark propagators of the proton should start from the same site on the source grid.
In this case, we can drop the terms with $\sum_{i \neq j}$ and $\sum_{i \neq j \, or \, j \neq k \, or \, k \neq i}$ which are all contamination without useful signal.
For the pure high mode contribution and mixed $G^H$ and $G^L$, this is implemented by the $Z_3$ noise.  For the case with all low modes, we can implement it by hand, since we have the wavefunctions in $G^L$.
\begin{eqnarray}
C(G, G, G) =& C(G^H, G^H, G^H) + \sum_{i} C(\theta_i G^L_i, \theta_i G^L_i, \theta_i G^L_i) \nonumber \\
           &+ \sum_{i} G(\theta_i G^L_i, G^H, G^H) + \sum_{i} G(G^H, \theta_i G^L_i, G^H) + \sum_{i} G(G^H, G^H, \theta_i G^L_i) \nonumber \\
           &+ \sum_{i} G(\theta_i G^L_i, \theta_i G^L_i, G^H) + \sum_{i} G(\theta_i G^L_i, G^H, \theta_i G^L_i) + \sum_{i} G(G^H, \theta_i G^L_i, \theta_i G^L_i) \nonumber \\
           =& G(G^H+\sum_{i} \theta_i G^L_i, G^H, G^H) \nonumber \\
           &+ G(G^H, \sum_{i} \theta_i G^L_i, G^H) + G(G^H, G^H, \sum_{i} \theta_i G^L_i) \nonumber \\
           &+ \sum_{i} \left[ G(\theta_i G^L_i, \theta_i G^L_i, G^H) + G(\theta_i G^L_i, G^H, \theta_i G^L_i) + G(G^H, \theta_i G^L_i, \theta_i G^L_i)  \right. \nonumber \\
           &+ \left. G(\theta_i G^L_i, \theta_i G^L_i, \theta_i G^L_i) \right]
\end{eqnarray}

With this low mode substitution technique (LMS)~\cite{lms1,lms2}, the proton correlation function acquires only errors from noises on the high modes and mixed $G^H$ and $G^L$.
Similarly, one can use the low mode substitution technique on propagators from smeared sources. The excited states are significantly suppressed by the smearing technique and the signal-noise ratio is much better.

\begin{figure}
\begin{center}
\includegraphics[width=3in]{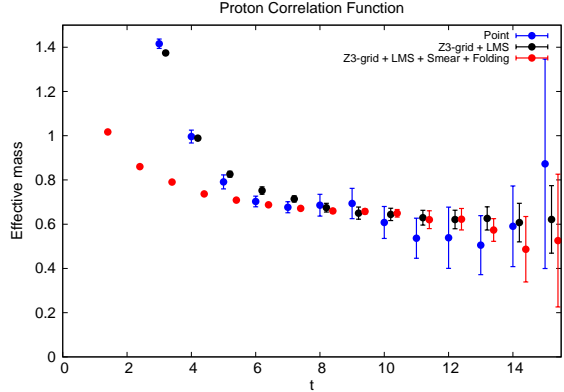}
\end{center}
\caption{
\label{fig:proton}
The comparison of proton effective masses. The fitting results for proton mass are $1.13(14)$GeV for point source, $1.08(5)$GeV for LMS with grid noise source and $1.14(2)$GeV for LMS on smeared grid source with folding.
}
\end{figure}

\subsection{The low mode average technique for the loop}

We use a similar technique to calculate the loop.
To get more statistics, we use the low mode average technique, wherein we use the low mode loop on all the space-time sites instead of on just the grid sites.

The loop for the scalar density is well saturated by low modes with light quark masses, but it is worse in the strange region and does not saturate at all in the charm region.

\begin{figure}
\begin{center}
\includegraphics[width=4in]{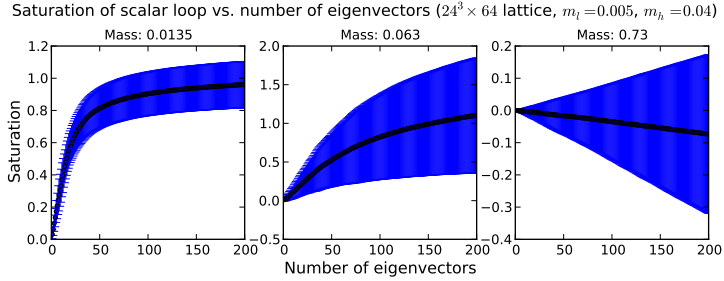}
\end{center}
\caption{
\label{fig:loop}
The low mode contributions to the loop.
}
\end{figure}

\section{Simulation details}

We use the overlap valence fermion on the $2+1$ flavor domain wall fermion configurations from RBC.
The lattice size is $24^3\times 64$ with $1/a = 1.73(3)$ GeV and the sea masses are $m_l=0.005$ and $m_h=0.04$.
We use 37 configurations for the calculation.

We use a double grid source for the nucleon two-point functions.
The sources are on the time slices $t=0$ and $t=32$ with $Z_3$ noises.
The grid cell is $(6,6,6)$ which means there are $(24/6)^3=64$ grid sites on one time slice.

We use partial dilution sources for the loop propagator.
The time slices are diluted to 4 pieces and an extra even-odd dilution is adopted.
Each diluted piece is a even-odd grid source with $Z_4$ noises.
The grid cell is $(4,4,4,2)$ on 4D space-time which means $(24/4)^3\times (64/2) /2=3456$ grid sites are inverted simultaneously as sources.

The strangeness with $m_{ud}=0.016$ (which gives $m_{\pi}=330\,{\rm MeV}$) for the nucleon and $m_s=0.067$ for the strange quark shows $ f_{T_{s}} = 0.048(15)$.
The contribution from the low modes and the high modes of the loops are $ {f_{T_{s}}}^L = 0.041(12)$ and $ {f_{T_{s}}}^H = 0.003(5)$, which shows the strangeness is dominated by the low modes.

The charmness with $m_c=0.63$ and $m_{ud}=0.016$ shows $f_{T_{c}} = 0.029(43)$.
The contribution from the low modes and the high modes of the loops are $ {f_{T_{c}}}^L = 0.042(12) $ and $ {f_{T_{c}}}^H = -0.016(45)$.

Since we are using the multi-mass algorithm to calculate the propagators, the form factors for light quarks are calculated simultaneously.
With both sea and valence $m_{ud}=0.016$, the disconnected insertion contribution to the $\pi N$ $\sigma$ term is $46(19)\,{\rm MeV}$.


\begin{figure}
\begin{center}
\includegraphics[width=0.45\hsize]{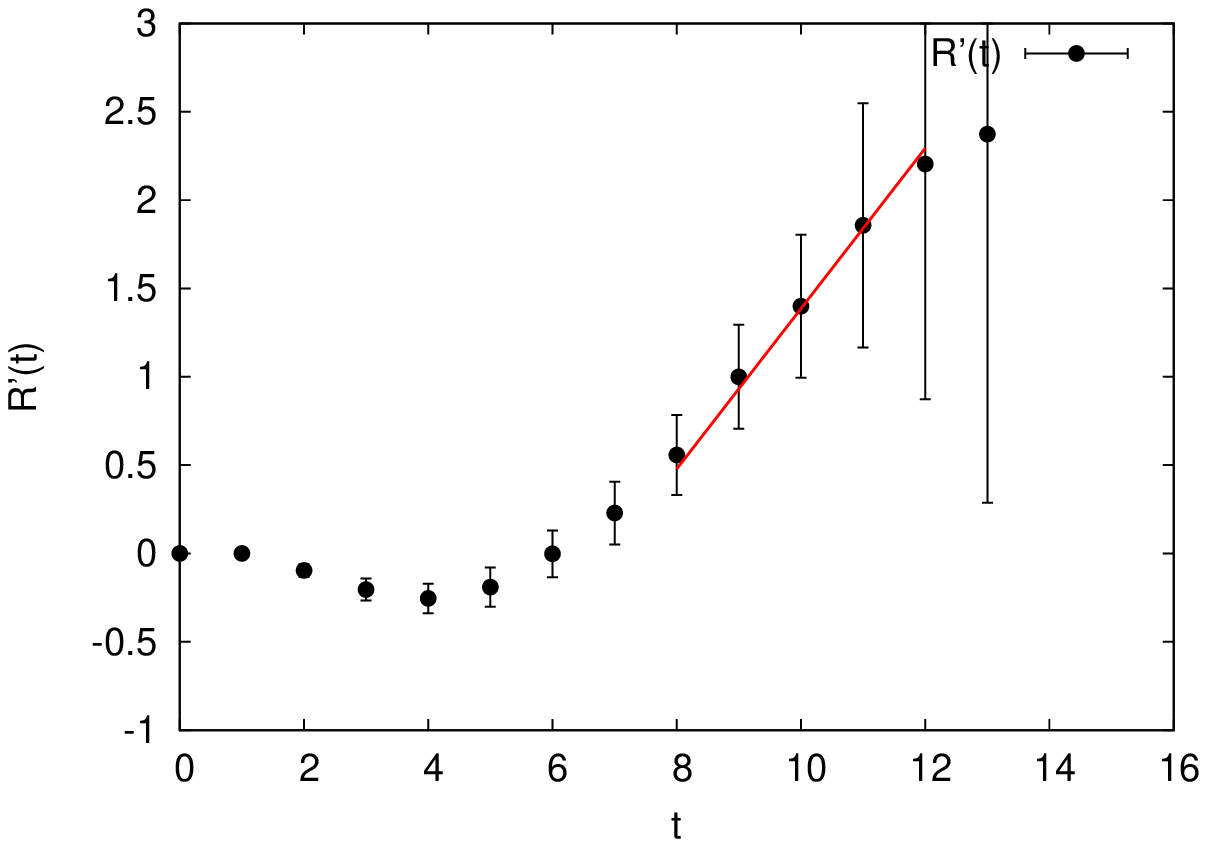}
\includegraphics[width=0.45\hsize]{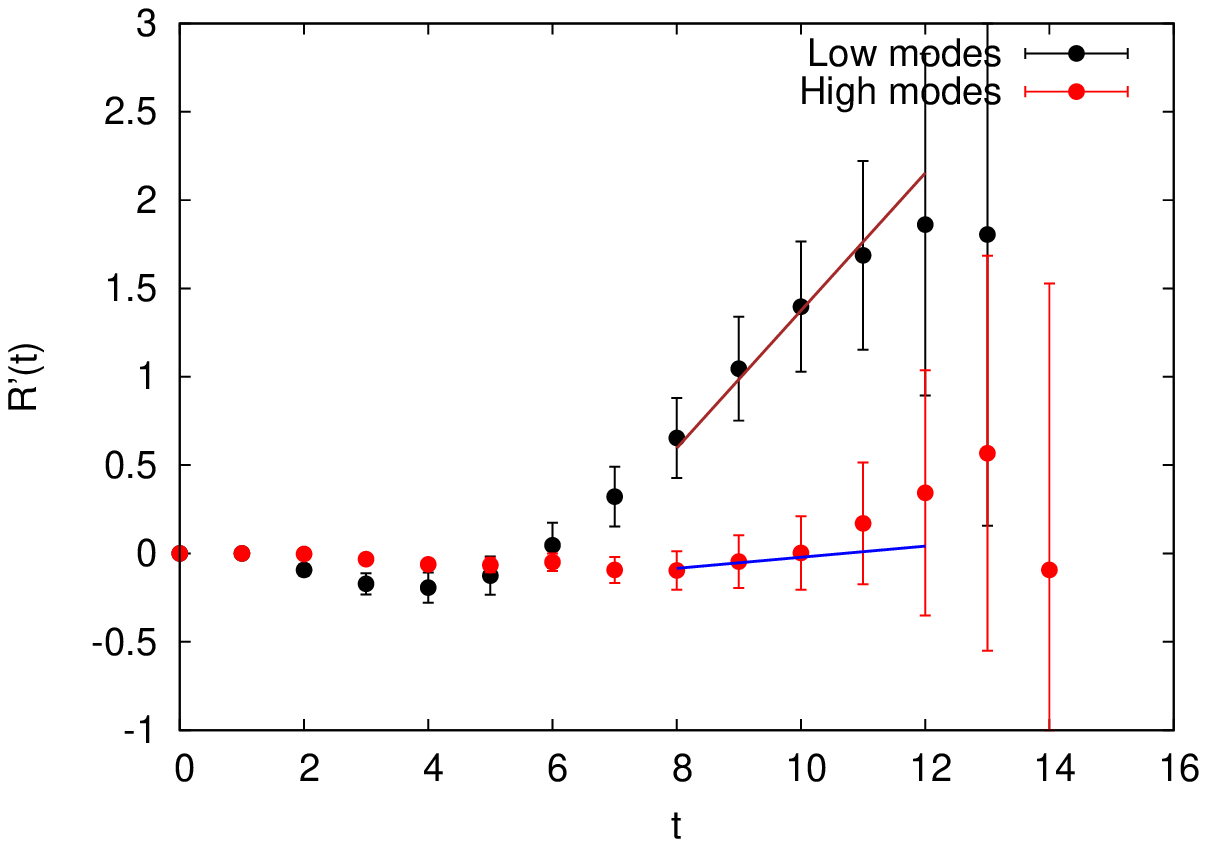}
\end{center}
\caption{
\label{fig:strangeness}
Left pane: the summed ratio for the strangeness content of nucleon.
Right pane: the contribution from the low modes and the high modes of the loop to the summed ratio for the strangeness content of nucleon.
}
\end{figure}


\begin{figure}
\begin{center}
\includegraphics[width=0.45\hsize]{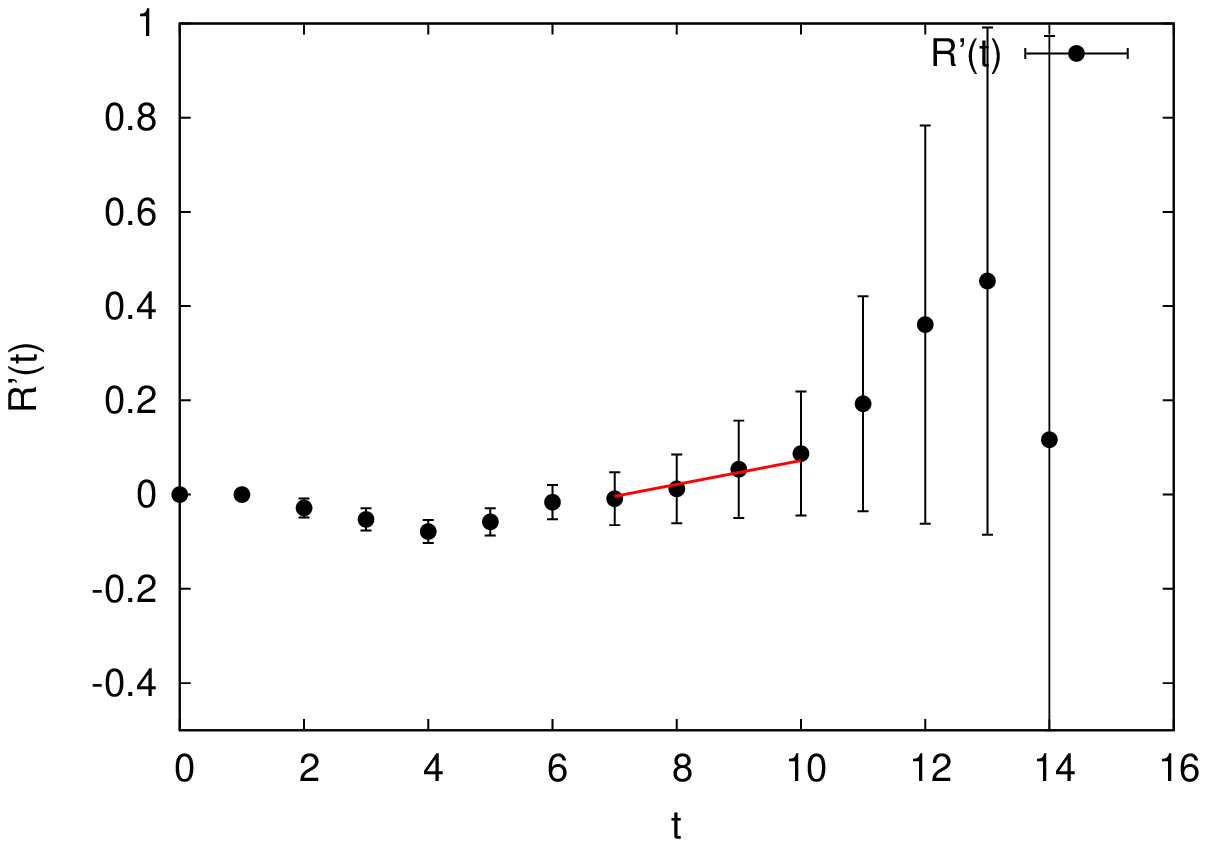}
\includegraphics[width=0.45\hsize]{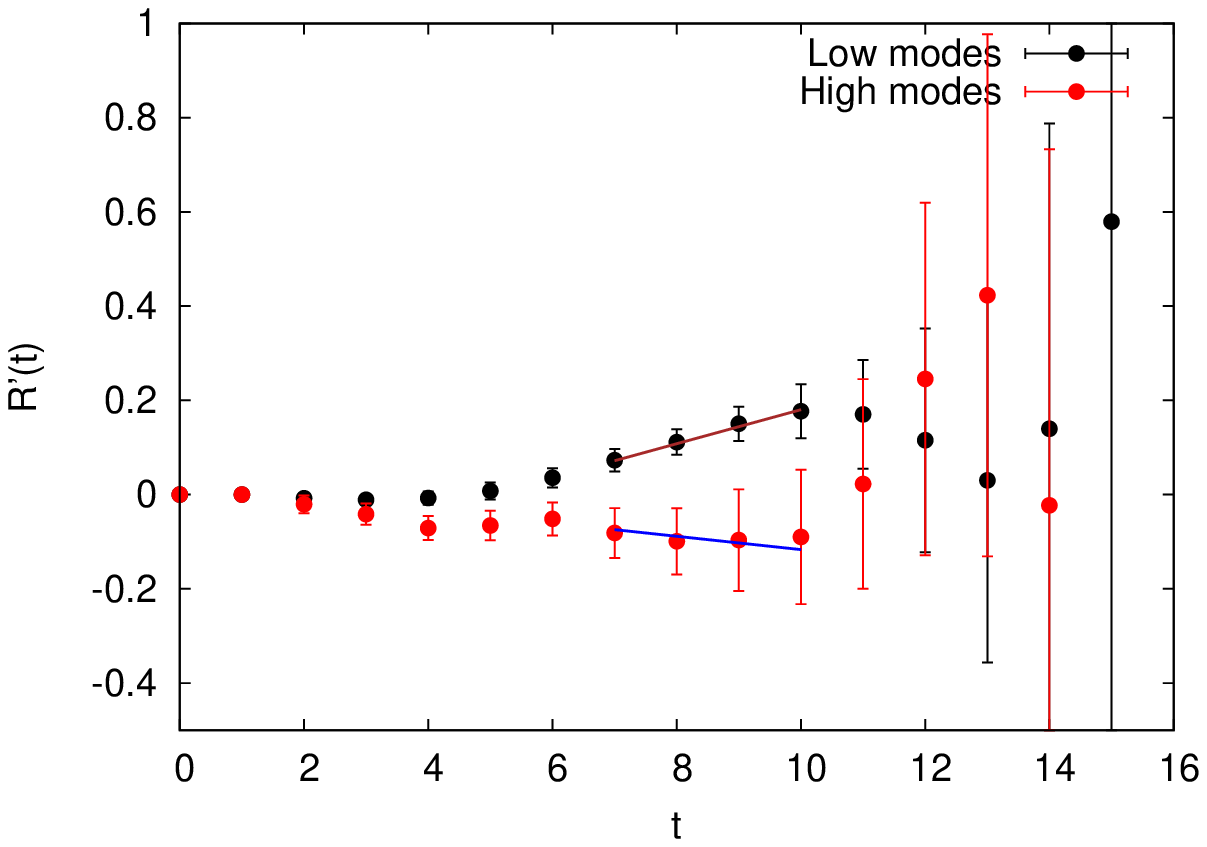}
\end{center}
\caption{
\label{fig:charmness}
Left pane: the summed ratio for the charmness content of nucleon.
Right pane: the contribution from the low modes and the high modes of the loop to the summed ratio for the charmness content of nucleon.
}
\end{figure}

\begin{figure}
\begin{center}
\includegraphics[width=3in]{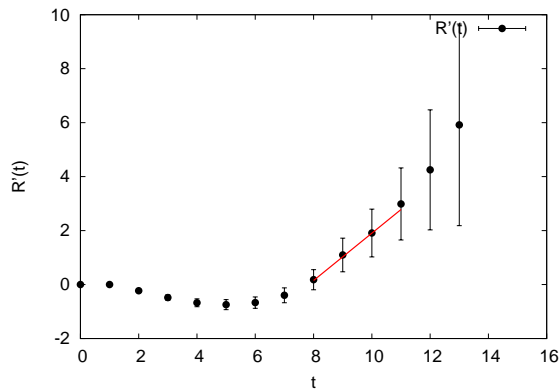}
\end{center}
\caption{
\label{fig:pinsigma}
The summed ratio for the disconnected insertion contribution to $\pi N$ $\sigma$ term.
}
\end{figure}

\section{Conclusions}

The strangeness and the charmness contents of nucleon and the $\pi N$ $\sigma$ term are presented.
The low mode substitution and the low mode average techniques are very helpful since the low modes play an important role for the scalar form factors.
For the nucleon valence quark which corresponds to $m_{\pi}=330\,{\rm MeV}$, we find a three-sigma signal for the strangeness content of the nucleon, $ f_{T_{s}} = 0.048(15)$,
but less than a one-sigma signal for the charmness content where the culprit is the noisy high-frequency modes.

This work is still preliminary with few configurations in one ensemble.
We will proceed with more configurations on both $24^3\times 64$ and $32^3\times 64$ lattices with different sea masses.
The variation method can be adopted to reduce the error bars further.

\end{document}